\documentclass[conference]{IEEEtran}

\usepackage{cite}

\ifCLASSINFOpdf
  \usepackage[pdftex]{graphicx}
  \pdfoutput=1
  \graphicspath{{./figs/}{./pics/}}
  \DeclareGraphicsExtensions{.pdf,.jpeg,.png}
  \usepackage{svg}
  \svgpath{{./figs/}}
\else
  \usepackage[dvips]{graphicx}
  \graphicspath{{./eps/}}
  \DeclareGraphicsExtensions{.eps}
  \svgpath{{./figs/}}
\fi
\ifCLASSOPTIONcompsoc
\usepackage[caption=false, font=normalsize, labelfont=sf, textfont=sf]{subfig}
\else
\usepackage[caption=false, font=footnotesize]{subfig}
\fi

\usepackage{amsmath}
\usepackage{mathtools}
\usepackage{tikz}

\newcommand*\circled[1]{\raisebox{.5pt}{\textcircled{\raisebox{-.9pt} {#1}}}}
\DeclareMathOperator*{\argmax}{arg\,max}

\newcounter{algorithm}

\usepackage{algorithmic}

\usepackage{array}

\usepackage{multicol}

\ifCLASSOPTIONcompsoc
  \usepackage[caption=false,font=normalsize,labelfont=sf,textfont=sf]{subfig}
\else
  \usepackage[caption=false,font=footnotesize]{subfig}
\fi
\usepackage{url}

\newif\ifpreprint
\preprinttrue 

\ifpreprint
	\usepackage[pdftex,%
    colorlinks=false,%
    linktocpage=true,%
    pdfstartpage=2,%
    pdfstartview=FitV,%
    breaklinks=true,%
    pdfpagemode=UseNone,%
    pageanchor=true,%
    pdfpagemode=UseOutlines,%
    plainpages=false,%
    bookmarksnumbered,%
    bookmarksopen=true,%
    bookmarksopenlevel=1,%
    hypertexnames=true,%
    pdfhighlight=/O,%
	hidelinks,%
    breaklinks,%
    pdftitle={Handover Optimality in Heterogeneous Networks},%
    pdfauthor={Lorenzo Di Gregorio},%
    pdfsubject={IEEE 5G World Forum 2019},%
    pdfkeywords={Gittins Index, Linear-Fractional Programming, Handover, Network Selection},%
	]{hyperref}
	\usepackage{fancyhdr}
\fi

\begin{document}

\ifpreprint
\begin{titlepage}
	
	\begin{center}
		{\Huge\bf Disclaimer and Legal Information}
	\end{center}

{\Large

\vfill

\begin{center}
	{\bf Preamble}
\end{center}

This document is a preprint paper accepted by the IEEE 5G World Forum 2019.  Its disclosure has been approved by Intel Corporation.

\vfill

\begin{center}
	{\bf Use of this material}
\end{center}

\textcopyright 2019 IEEE.  Personal use of this material is permitted.  Permission from IEEE must be obtained for all other uses, in any current or future media, including reprinting/republishing this material for advertising or promotional purposes, creating new collective works, for resale or redistribution to servers or lists, or reuse of any copyrighted component of this work in other works

\vfill

\begin{center}
	{\bf Disclaimer}
\end{center}

All opinions expressed in this document are those of the authors individually and are
not reflective or indicative of the opinions and positions of the authors' employer.

The technology described in this document is or could be under development and is being presented solely for the purpose of soliciting feedback. The content and any information in this presentation shall in no way be regarded as a warranty or guarantee of conditions of characteristics.

This document reflects the current state of the subject matter and may unilaterally be changed by Intel Corporation and/or its affiliated companies (hereinafter referred to as ``Intel'') at any time. Unless otherwise formally agreed with Intel, Intel assumes no warranties or liabilities of any kind, including without limitation warranties of non-infringement of intellectual property rights of any third party with respect to the content and information given in this document.

\vfill
}

\end{titlepage}
\fi

%
\title{Handover Optimality in Heterogeneous Networks}

\author{\IEEEauthorblockN{Lorenzo Di Gregorio, Valerio Frascolla}
\IEEEauthorblockA{Intel Deutschland GmbH\\
email: lorenzo.di.gregorio@intel.com, valerio.frascolla@intel.com}}


%


\maketitle

\ifpreprint
	\thispagestyle{fancy}
\fi

\begin{abstract}
This paper introduces a new theoretical framework for optimal handover procedures in heterogeneous networks by devising the novel fractional Gittins indices, which are dynamical priorities whose values can be statically associated to the decision states of evolving processes representing handover alternatives.  The simple policy of activating at any time the one process currently at highest priority optimizes the bandwidth of a handover, if all other inactive processes remain idle.  However, numerical evidence shows that in practice this condition can be relaxed for a wide range of handover models, because the bandwidth actually achieved by the policy never deviates for more than 12\% from the optimally achievable bandwidth and remains in median within a deviation of 2\% from this optimum.
\end{abstract}


%
\IEEEpeerreviewmaketitle

\section{Introduction}\label{sec:Introduction}
This paper refers with {\em handover} to the procedure of transferring one communication session from one physical {\em or} logical channel to another.
With the ongoing industrial evolution \cite{advanc2014} toward access proliferation and densification driven by new IoT and 5G services, as described in \cite{DBLP:conf/camad/2018}, handover procedures need to increase efficiency significantly, as they are expected to face much more challenging conditions than ever before. For example, in vehicular networking users might trigger a handover every few seconds, demanding a very good allocation likelihood (e.g. \cite{DBLP:journals/tvt/VasudevaSLG17}) as well as a timely switch of the downlink data path from the serving gateway to the target node (e.g. \cite{Boukerche:2017:MIH:3022634.2996451}).

Although handover functionalities in next generation heterogeneous networks, i.e. networks under inhomogeneous wireless access technologies and protocols, shall be widely diverse and are to a large extent still to be defined, their predictive control structure is usually modeled by means of a Markovian framework.  This paper analyzes optimality within that framework and shows how a surprisingly simple policy of {\em associating static priorities to decision stages of handover alternative processes} can induce a {\em near-optimal} handover decision, with general validity on generic handover procedures.

This paper formulates the handover decision problem as {\em multiarmed bandit} \cite{DBLP:conf/wcnc/ShenS17, a11020013}, which, although in its most general formulation (said ``{\em restless}'') is a too complex problem (PSPACE complexity), work in \cite{gittins_2011} has proven that the policy of always selecting the process (said ``arm'') whose current state bears the highest statically associated priority (Gittins index) maximizes the {\em total accrual of one discounted additive utility} under some relatively mild conditions, given in \cite{Nino-Mora2007,RePEc:spr:topjnl:v:15:y:2007:i:2:p:217-219}. Such {\em index policy} was shown to accrue a {\em near-optimal} lower bound in the general restless case obtained relaxing the said conditions (see \cite{10.2307/3214547,4b0388e8899343dcabb7dfa22f6070a9}). Efficient algorithms for calculating Gittins indices have been devised for Markov chains \cite{chakravorty2014a}.

However, handover procedures do not fit to models for calculating Gittins indices because they are not memoryless and the objective of their optimization is not the maximization of one additive utility. Available literature {\em stretches} formulations by lumping memory into exploding state spaces and either deferring the question of what the utility shall be or introducing heuristic costs like adjustments of signal-to-noise ratios, which are not additive and deceive optimization.

The novelty presented by this paper overcomes the aforementioned limitations by formulating a ratio maximization problem and deriving a novel {\em fractional} Gittins index to obtain priorities for {\em semi-Markov} processes, consisting of Markov chains {\em embedded} in decision stages of stochastic processes.  Under this approach, state transitions of handover procedures can be directly represented as embedded Markov chains and related priorities can be statically associated to decision stages of individual handover alternatives.  A computationally lightweight algorithm is given to calculate these priorities introducing {\em sojourn times per state as a second utility} in the stopping problem resolved in \cite{SONIN20081526}.

The contribution of this paper is twofold: Gittins indices are ported to the semi-Markovian case for ratio optimization and techniques based on decision thresholds are shown to conceptually scale for future networks: one algorithm to operatively calculate such thresholds and one numerical analysis are given.

\section{Related Work}

This work demonstrates the scalability of policies based decision thresholds, which are also known in networking communities as ``sticky'' policies.  Starting with LTE release 10, a sticky handover procedure for access networks consisting of low and high power nodes, has been defined in terms of cell range extension (CRE).  CRE consists of adding a bias onto the signal strength measurements received by a low power node.  This bias must be devised to pitch a threshold for triggering a handover, so that it trades off throughput due to weaker signal of a low power node against congestion in accessing a high power node.  This work shows that, while such a simple scheme is insufficient in the general case, stochastic accumulation toward decision thresholds is sufficient and scalable for achieving near-optimal decisions.

Vehicular networks are subject to higher handover failure rates (see \cite{DBLP:journals/tvt/VasudevaSLG17}).  However, the problem of {\em frequent handover} (e.g. \cite{DBLP:conf/wcnc/ShenS17}) is technically ill-posed: the handover is not ``too'' frequent, it is excessively suboptimal and the proper frequency shall be achieved if a near-optimal predictive policy is installed.  In this paper, no distinction is made between network layers: {\em predictive handover} refers also to handover procedures in upper layers (see \cite{Boukerche:2017:MIH:3022634.2996451}), as long as they can be regarded as embedding semi-Markov processes for control.

Optimal handovers as solution of a Markov decision processes have been pursued in \cite{DBLP:conf/eucnc/MezzavillaGPRZ16}.  However, that work presents the policy optimization of discrete-time Markov processes on a cartesian product of Markovian state spaces with rates as rewards.  Unfortunately, that approach is affected by a number of well-known and already mentioned weaknesses (e.g. non-additive utilities, state-space explosion, exponential sojourn times ...) and indeed the Gittins indices have been developed exactly to prevent the state-space explosion in finding optimal policies on a cartesian product of markovian state spaces.

The classical {\em uniformization} only transforms continuous-time Markov chains with Poisson-distributed sojourn times into discrete ones.  While literature exists on multi-objective multiarmed bandits (e.g. \cite{6707036,7526494}), authors employ a {\em scalarization}, which optimizes accrual and cannot induce optimality on a semi-Markov process.  Contrarily, the solution presented in this paper achieves optimality by building on the convergence of a semi-Markov reward process, e.g. as shown in \cite{Sladky2005}.

The classical {\em discount} factor can be interpreted as probability of {\em termination} of a process: both {\em restart-in-state} \cite{Katehakis:1987:MBP:34030.34037} and {\em elimination} \cite{SONIN20081526} algorithms enable a more general state-dependent termination to be accounted into the calculation of the indices.  States of the semi-Markov processes considered in this work may include state-dependent transitions to one non-regenerative {\em termination} state: this is important for example to model likelihoods of handover failures.

Since the restless bandit is PSPACE-hard \cite{315792}, for large models in general the only feasible solution is the restful bandit one.  The claim that this solution is near-optimal for the restless bandit is a Lagrangian result based on reasoning in \cite{whittle_1988, ninomora2001}, which explains that {\em some} indices, not necessarily the ones of the corresponding restful bandit, are optimal or asymptotically near-optimal.  However, the indices are an efficient heuristic \cite{10.2307/3214547}, and the ones for the restful case  converge to optimality as passive activity decreases.  Section \ref{sec:Results} presents a numerical analysis of this efficiency for {\em fractional} indices in the noteworthy restless case of the {\em switching bandit} \cite{RePEc:ecm:emetrp:v:62:y:1994:i:3:p:687-94}.

\section{Theory}\label{sec:Theory}

This paper makes no specific assumptions on handover procedures and regards every handover alternative as a semi-Markov process, whose reward per state is the amount of information (payload data) received during sojourn in that state.  The handover procedure is the realization of a selection among several alternative processes, each representing one evolution when {\em active}, i.e. selected as the one to carry out the handover, or another evolution when {\em passive}, i.e. the environment might change but the alternative is not selected.

Formally, every such process $i$ is described by an embedded Markov chain characterized by a state transition matrix $\bar{P}^{(i)}$ with state rewards $R^{(i)}$ and sojourn times $D^{(i)}$.  In literature, it is common to introduce a {\em discount factor} $\beta$ or {\em termination probability} $1-\beta$ so that $\bar{P}^{(i)}\doteq \textup{diag}(\beta^{(i)})P^{(i)}$, with $P^{(i)}$ as stochastic matrix.  With this notation, process $i$ is described by the a pair of chains $\{\bar{P}_{(\mathcal{A})}^{(i)}, R_{(\mathcal{A})}^{(i)}, D_{(\mathcal{A})}^{(i)}\}$ (active) and $\{\bar{P}_{(\mathcal{P})}^{(i)}, R_{(\mathcal{P})}^{(i)}, D_{(\mathcal{P})}^{(i)}\}$ (passive), among which a {\em policy} switches when selecting active and passive processes.

Obviously, the memory for rewards $R^{(i)}$ and intervals $D^{(i)}$ is still lumped into states of $P^{(i)}$, but the semi-Markov model relaxes the infeasible assumption of continuous-time Markov chains that sojourn times must be exponentially distributed, so only the actual states of the functionality are modeled and no additional states must be introduces to model sojourn times.

\subsection{Example}\label{sec:example}

\begin{figure}[!t]
	\centering
	{\footnotesize\includesvg[width=0.35\textwidth]{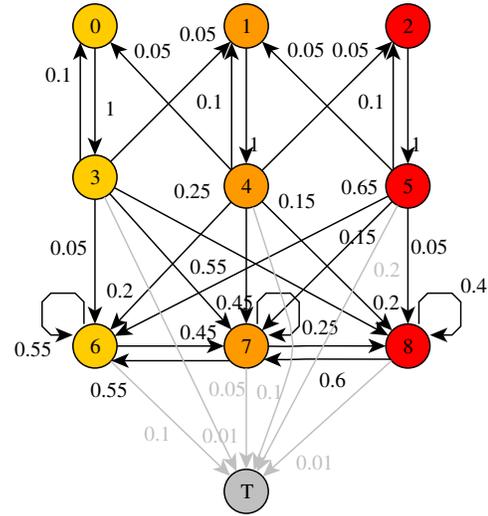}}
	\caption{Simplistic example: the received signal is weak in \circled{0}, \circled{3}, \circled{6}, medium in \circled{1}, \circled{4}, \circled{7} and strong in \circled{2}, \circled{5}, \circled{8}.  Handover is entered in \circled{0}, \circled{1}, \circled{2} and exited in \circled{6}, \circled{7}, \circled{8} which represent an established link with fluctuating signal strength.  Ongoing handover with backoffs takes place in \circled{3}, \circled{4}, \circled{5}.}
	\label{fig:handover_model}
\end{figure}

The application of the theory of section \ref{sec:Theory} is illustrated in figure \ref{fig:handover_model}.  The transitions among \circled{6}, \circled{7} and \circled{8} model a link whose received signal fluctuates between weak, medium and strong strengths.  The remaining transitions model a handover under fluctuations.  The alternatives consist of retaining the link under \circled{6}, \circled{7} or \circled{8}, or attempting a handover by transitioning to \circled{0}, \circled{1} or \circled{2}, depending on the received weak, medium or strong signal.  Transitions to \circled{T} model connection loss probabilities. The graph in figure \ref{fig:handover_model} is given in matrix form in (\ref{eq:handover_model}) and the link model is given by the submatrix $P(6:8,6:8)$ with survival probabilities $\beta(6:8)$.

\begin{equation}\label{eq:handover_model}
P=\begin{bsmallmatrix*}[l]
0    & 0    & 0    & 1    & 0    & 0    & 0    & 0    & 0    \\
0    & 0    & 0    & 0    & 1    & 0    & 0    & 0    & 0    \\
0    & 0    & 0    & 0    & 0    & 1    & 0    & 0    & 0    \\
0.1  & 0.05 & 0    & 0    & 0    & 0    & 0.05 & 0.55 & 0.25 \\
0.05 & 0.1  & 0.05 & 0    & 0    & 0    & 0.2  & 0.45 & 0.15 \\
0    & 0.05 & 0.1  & 0    & 0    & 0    & 0.65 & 0.15 & 0.05 \\
0    & 0    & 0    & 0    & 0    & 0    & 0.55 & 0.45 & 0    \\
0    & 0    & 0    & 0    & 0    & 0    & 0.55 & 0.25 & 0.2  \\
0    & 0    & 0    & 0    & 0    & 0    & 0    & 0.6  & 0.4
\end{bsmallmatrix*}
\beta=\begin{bsmallmatrix*}[l]
1    \\
1    \\
1    \\
0.99 \\
0.9  \\
0.8  \\
0.9  \\
0.95 \\
0.99
\end{bsmallmatrix*}
\end{equation}

The actual stochastic transition matrix of the Markov chain in figure \ref{fig:handover_model} is
$\begin{bsmallmatrix}
\textup{diag}(\beta)P & \vec{1}-\beta \\
\vec{0}           & 1
\end{bsmallmatrix}$ with \circled{T} as absorbing state.

We assign average sojourn times per state as $D = \begin{bsmallmatrix}0.1 & 0.1 & 0.1 & 1 & 1 & 1 & 5 & 1 & 2.5 \end{bsmallmatrix}^T$ and average amount of information received per state during sojourn as $R = \begin{bsmallmatrix}0 & 0 & 0 & 0 & 0 & 0 & 0.1 & 1 & 10 \end{bsmallmatrix}^T$.

For compactness of exposition, in this bandit problem the active alternatives are the chain $\{\bar{P}, R, D\}$ and its closed component $\{\bar{P}(6:8,6:8), R(6:8), D(6:8)\}$, the passive ones are implied to be identity matrices with zero reward and unitary sojourn times.  Since $P$ is not irreducible, the indices are calculated by algorithm 1 in row 4 employing \eqref{eq:lfp_as_lp} rather than \eqref{eq:policy_iteration}. A variation of \eqref{eq:policy_iteration} would have been also possible.

After calculations and truncations for readability, the indices for {\em retaining} the link (\circled{6}, \circled{7} and \circled{8}) turn out to be $\begin{bsmallmatrix}0.48 & 2.32 & 4 \end{bsmallmatrix}$.  They must be compared against the indices for {\em initiating} a handover (\circled{0}, \circled{1} and \circled{2}), which are $\begin{bsmallmatrix}2.13 & 1.85 & 1.09\end{bsmallmatrix}$.  Unsurprisingly, we see that only on a weak signal state (\circled{6} at 0.48) it is optimal to initiate a handover (\circled{0} at 2.13) and carry it on (\circled{3}, \circled{4} and \circled{5} at $\begin{bsmallmatrix}2.18 & 1.90 & 1.15 \end{bsmallmatrix}$) as long as the signal remains weak (\circled{3} at 2.18635297).

\begin{figure}[!t]
\centering
\includesvg[width=0.45\textwidth]{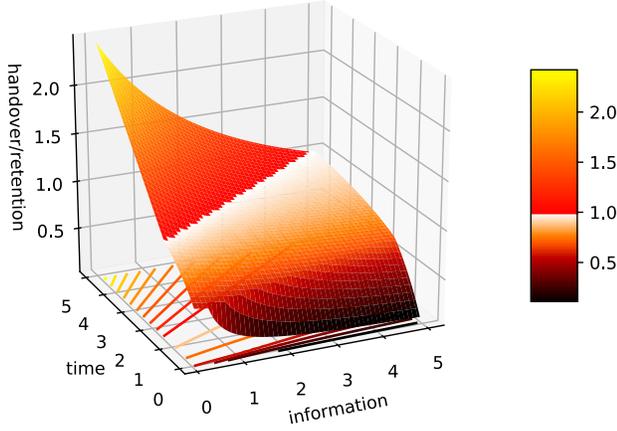}
\caption{The region where the handover/retention ratio is grater than 1 is the optimal one to initiate a new handover rather than to retain the current link.}
\label{fig:handover_example}
\end{figure}

The related question of which middle signal strength should trigger a handover is dealt with in figure \ref{fig:handover_example}, showing the ratio of the index of \circled{1} against the index of \circled{7} as we sweep $R(7)$ (information) and $D(7)$ (time): where the ratio grows above 1, the index of \circled{1} is greater than the index of \circled{7} and it is optimal to initiate a handover rather than retain the link.

\subsection{Linear-Fractional Markov Decision Problems}

This short paper can only present bare facts and rationales for the actual calculations of optimality, delegating proofs to a larger work and referencing \cite{Aggarwal1977} for a foundation.

It can be proved that the optimal bandwidth maximization policy $\pi$ among $a$ alternatives $\{\bar{P}^{(i)},R^{(i)},D^{(i)}\}, \forall i \in 1,\ldots,a$, consisting of selecting alternative $\pi(i)$ when in state $i$, is given by solving the linear fractional program in \eqref{eq:lfp_bw_maximization} to calculate the row vectors $x_1,\ldots,x_a$ from which $\pi$ is obtained, with $\alpha: \sum_j\alpha_j=1$ as the initial state probability distribution. This ratio maximization can be solved as linear program through the Charnes-Cooper transformation of \eqref{eq:lfp} in \eqref{eq:lfp_as_lp}. The solution of \eqref{eq:lfp_as_lp} with $\eta = 0$ and $\theta = 0$, which is the case in \eqref{eq:lfp_bw_maximization}, can be shown to be determined by {\em policy iteration} \eqref{eq:policy_iteration} for {\em irreducible} Markov chains.  A solution as policy iteration for non-irreducible chains is possible but more complicated. 
\begin{equation}\label{eq:lfp_bw_maximization}
\begin{gathered}
\max_{x}\quad\frac{[R^{(1)},\ldots,R^{(a)}]\cdot[x_1,\ldots,x_a]^T}{[D^{(1)},\ldots,D^{(a)}]\cdot[x_1,\ldots,x_a]^T}\\
\textup{s.t.}\quad [I-(\bar{P}^{(1)})^T,\ldots,I-(\bar{P}^{(a)})^T]\cdot\begin{bsmallmatrix}
   x_1^T \\
   \vdots\\
   x_a^T\end{bsmallmatrix}=\alpha\\
\pi = \argmax [x_1^T,\ldots,x_a^T]
\end{gathered}
\end{equation}

\noindent\begin{minipage}[t]{.45\linewidth}
	\begin{equation}\label{eq:lfp}
	\begin{gathered}
	\min_{x}\quad \frac{\bar{R} x + \eta}{\bar{D} x + \theta} \\
	\textup{s.t.}\quad Ax \leq  \alpha \\
	{\small\begin{aligned}
	\bar{R}&\doteq [R^{(1)},\ldots,R^{(a)}] \\
	\bar{D}&\doteq [D^{(1)},\ldots,D^{(a)}] \\
	     x &\doteq [x_1,\ldots,x_a] \\
	\end{aligned}}
	\end{gathered}
	\end{equation}
\end{minipage}%
\begin{minipage}[t]{.55\linewidth}
	\begin{equation}\label{eq:lfp_as_lp}
	\begin{gathered}
	\min_{y, \gamma}\quad [\bar{R}, \eta]
		\begin{bmatrix}
		y \\
		\gamma  \\
		\end{bmatrix}, x=y/\gamma\\
	\begin{aligned}
	\textup{s.t.}\quad
		\begin{bmatrix}
		A       & -\alpha \\
		\vec{0} & -1 \\
		\end{bmatrix}
			\begin{bmatrix}
			y \\
			\gamma \\
			\end{bmatrix} &\leq  0\\
		[\bar{D}, \theta]
			\begin{bmatrix}
			y \\
			\gamma  \\
			\end{bmatrix}&= 1\\
	\end{aligned}
	\end{gathered}
	\end{equation}
\end{minipage}\\
\smallskip

Say $n$ the space state dimension ($P^{(i)}\in\mathbf{R}^{n\times n}, \forall i$), then $s$ is the solution of the dual of \eqref{eq:lfp_as_lp} determined starting the {\em policy evaluation} with $s=0$ in \eqref{eq:policy_evaluation} and iterating over {\em value determination} in \eqref{eq:value_determination} with $i+1 \rightarrow i$ until \eqref{eq:policy_evaluation} returns $\pi^{(i+1)}=\pi^{(i)}$ for some $i$.  This policy is $\pi$ sought with \eqref{eq:lfp_bw_maximization}.

\begin{subequations}\label{eq:policy_iteration}
			\begin{equation}\label{eq:policy_evaluation}
			\begin{gathered}
			\pi^{(i)} = \argmax [t^{(1)},\ldots,t^{(a)}] \\
			t^{(i)}\doteq [(R^{(i)})^T-(I-\bar{P}^{(i)})s]/(D^{(i)})^T
			\end{gathered}
			\end{equation}

			\begin{equation}\label{eq:value_determination}
			\begin{gathered}
			\begin{bmatrix}
			I-\bar{P}_{\pi^{(i)}} & D_{\pi^{(i)}} \\
			\alpha & 0
			\end{bmatrix}%
			\begin{bmatrix}
			s \\
			g
			\end{bmatrix}%
			=%
			\begin{bmatrix}
			R_{\pi^{(i)}} \\
			0
			\end{bmatrix}\\
			\bar{P}_{\pi^{(i)}}\doteq
				\begin{bmatrix}
				\bar{P}^{\left(\pi^{(i)}(1)\right)}(1,1) & \cdots & \bar{P}^{\left(\pi^{(i)}(1)\right)}(1,n) \\
				\vdots                &        & \vdots \\
				\bar{P}^{\left(\pi^{(i)}(n)\right)}(n,1) & \cdots & \bar{P}^{\left(\pi^{(i)}(n)\right)}(n,n)
				\end{bmatrix}\\
			R_{\pi^{(i)}}\doteq[R^{\left(\pi^{(i)}(1)\right)},\ldots,R^{\left(\pi^{(i)}(a)\right)}]\\
			D_{\pi^{(i)}}\doteq[D^{\left(\pi^{(i)}(1)\right)},\ldots,D^{\left(\pi^{(i)}(a)\right)}]\\
			\end{gathered}
			\end{equation}
\end{subequations}

\subsection{Optimality}\label{sec:Optimality}

\begin{figure}[!t]
	\centering
	\includesvg[width=0.45\textwidth]{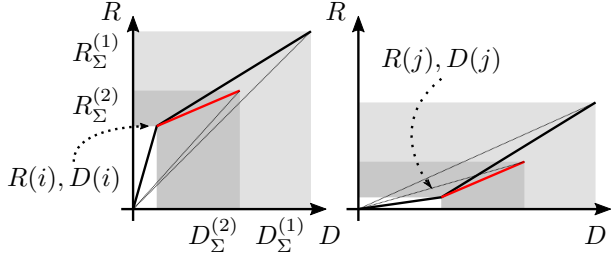}
	\caption{Example showing that a lower ratio $R^{(2)}_\Sigma/D^{(2)}_\Sigma$ can lead to a higher overall ratio $(R^{(2)}_\Sigma+R(j))/(D^{(2)}_\Sigma+D(j))$ depending on initial conditions.}
	\label{fig:ratio_optimality}
\end{figure}

A peculiarity of the {\em discounted Markov ratio decision process} is that $\pi$ is dependent on $\alpha$ through \eqref{eq:lfp_bw_maximization}, hence the initial state can affect the optimal policy.  This effect is shown in figure~\ref{fig:ratio_optimality}: a start in state $i$, accruing $R(i), D(i)$, induces the total discounted rewards of alternative 2, $R^{(2)}_\Sigma, D^{(2)}_\Sigma$, to deliver the maximum ratio $(R^{(2)}_\Sigma+R(i))/(D^{(2)}_\Sigma+D(i))$.  A start in state $j$, accruing $R(i), D(i)$, induces the total discounted rewards of alternative 1 to deliver the maximum ratio $(R^{(1)}_\Sigma+R(j))/(D^{(1)}_\Sigma+D(j))$.

The outcome of this effect in simple words is that, if one controller has achieved a very high bandwidth in the past, then it is optimal to stop quickly rather than endure at a lower bandwidth than in the past.

However, breaking up rather than continuing cannot be an optimality criterion in communications, hence this paper defines as optimal a policy which maximizes the {\em expected} bandwidth, in probabilistic terms, under the current state as initial one and disregards the past bandwidth achieved in the evolution up to the current state.  This definition matches the common sense by maximizing the ``future'' bandwidth and does not rule out ``investing'' into a transition through low bandwidth states to achieve a high bandwidth one, as it would be the case in a handover.

In a fully observable process, $\alpha$  has only one element at 1 and the others at 0, i.e. $\alpha$ is 1-sparse with $\sum_{i=1}^n \alpha(i)=1$.  For every state, the optimal policy must be calculated assuming that state as initial and $n$ optimal policy calculations for $n$ initial states are necessary to solve the decision problem.

\subsection{Multiarmed Bandit}\label{sec:Bandit}

A multiarmed bandit in a Markovian or semi-Markovian framework consists of a set of $m$ Markov or semi-Markov reward processes, with each process $i$ selecting among the two chains $\{\bar{P}_{(\mathcal{A})}^{(i)}, R_{(\mathcal{A})}^{(i)}, D_{(\mathcal{A})}^{(i)}\}$ and  $\{\bar{P}_{(\mathcal{P})}^{(i)}, R_{(\mathcal{P})}^{(i)}, D_{(\mathcal{P})}^{(i)}\}$.

In a conventional restful bandit, $\bar{P}_{(\mathcal{P})}=I$, $R_{(\mathcal{P})}=\vec{0}$ and only one out of $m$ processes is active at any time.  The sought-for solution is the policy that maximizes the accrued total reward of the bandit until termination.  Instead, as explained aboce, in this paper the {\em ratio} of the total {\em expected} reward over total {\em expected} time until termination shall be maximized.

\begin{equation}\label{eq:bandit_as_mdp}
\begin{gathered}
\begin{aligned}
\hat{P}^{(j)} & = \otimes_{i=1}^a \bar{P}^{(i)} \\
\hat{R}^{(j)} & = \sum_{i=1}^a \vec{1}_{\prod_{j=0}^{i-1} n_j \times 1} \otimes R^{(i)} \otimes \vec{1}_{\prod_{j=i+1}^{a+1} n_j \times 1}\\
\hat{D}^{(j)} & = \sum_{i=1}^a \vec{1}_{\prod_{j=0}^{i-1} n_j \times 1} \otimes D^{(i)} \otimes \vec{1}_{\prod_{j=i+1}^{a+1} n_j \times 1}\\
\textup{where} & \quad n_0\doteq 1, n_{a+1}\doteq 1, n_i: R^{(i)} \in \mathbf{R}^{n_i}
\end{aligned}\\
\bar{P}^{(i)}, \bar{R}^{(i)}, \bar{D}^{(i)}=
	\begin{cases}
	\bar{P}_{(\mathcal{A})}^{(i)}, R_{(\mathcal{A})}^{(i)}, D_{(\mathcal{A})}^{(i)}, & \text{if $i=j$}.\\
	\bar{P}_{(\mathcal{P})}^{(i)}, R_{(\mathcal{P})}^{(i)}, D_{(\mathcal{P})}^{(i)}, & \text{otherwise}.
	\end{cases}
\end{gathered}
\end{equation}

The multiarmed bandit problem can be solved formulating it as a conventional Markov decision process among $a$ alternatives $\{\hat{P}^{(j)}, \hat{R}^{(j)}, \hat{D}^{(j)}\}, \forall j=1,\ldots,a$ built by formula \eqref{eq:bandit_as_mdp} where $\otimes$ is the Kronecker product.

However, if for example all $\bar{P}^{(i)}$ are $n\times n$ matrices, the linear system in \eqref{eq:value_determination} must be solved $a^2$ times ($a$ alternatives in $a$ initial states) for $n^a+1$ equalities in $n^a+1$ variables: this upscaling faces rather soon a state space explosion.

\subsection{Gittins Indices}\label{sec:Gittins}

The Gittins indices are real scalar values associated to the states of $\bar{P}^{(i)}$ with the property that selecting the alternative whose current state presents the highest index maximizes the accrued total reward, so the Gittins indices constitute an optimal policy for the conventional multiarmed bandit and do not cause a state space explosion when upscaling the problem.

In this paper we derive a novel {\em fractional} Gittins index for ratio maximization rather than reward maximization.

Following the restart-in-state formulation \cite{Katehakis:1987:MBP:34030.34037}, the Gittins index of state $j$ is the maximum ratio achievable by a Markov decision process whose alternatives at every decision stage are to continue or restart in state $j$.  The optimal policy can be found by policy iteration \eqref{eq:policy_iteration}, or solving \eqref{eq:lfp} through \eqref{eq:lfp_as_lp}.  Algorithm 1 describes the calculation in details employing \eqref{eq:policy_iteration} and its distinguishing feature against other algorithms is that it calculates the Gittins index of one specific state $S$ without calculating all other Gittins indices associated to all other states of the process.  However, the convergence of \eqref{eq:policy_iteration} can be computationally intensive.

\begin{algorithmic}[1]
	\stepcounter{algorithm}
	\renewcommand{\algorithmicensure}{\textbf{Algorithm 1:}}
	\renewcommand{\algorithmicrequire}{\textbf{Input:}}
	\ENSURE \textbf{Fractional Gittins Index of Single State}\\
	\textit{Calculation by restart-in-state optimization}\label{alg:restart-in-state}
	\REQUIRE
		\begin{tabular}{cp{17.5em}}
			$\bar{P}^{(i)}_{(\mathcal{A})}$ & $\forall x,y:\bar{P}^{(i)}_{(\mathcal{A})}(x,y)$ is the probability that process $i$ in state $x$ transitions to state $y$. \\
			$R^{(i)}_{(\mathcal{A})}$       & $\forall x: R^{(i)}_{(\mathcal{A})}(x)$ is the average information transported by process $i$ in state $x$. \\
			$D^{(i)}_{(\mathcal{A})}$       & $\forall x: D^{(i)}_{(\mathcal{A})}(x)$ is the average time process $i$ has sojourned in state $x$. \\
			$S$                             & state for which the index is calculated. \\
		\end{tabular}
	\STATE $\bar{P}^{(1)}, R^{(1)}, D^{(1)} \leftarrow \bar{P}^{(i)}_{(\mathcal{A})}, R^{(i)}_{(\mathcal{A})}, D^{(i)}_{(\mathcal{A})}$
	\STATE $\bar{P}^{(2)} \leftarrow
		\begin{bsmallmatrix}
			\bar{P}^{(i)}_{(\mathcal{A})}(S,1) & \ldots & \bar{P}^{(i)}_{(\mathcal{A})}(S,n) \\
					\vdots                     &        &             \vdots \\
			\bar{P}^{(i)}_{(\mathcal{A})}(S,1) & \ldots & \bar{P}^{(i)}_{(\mathcal{A})}(S,n)
		\end{bsmallmatrix}$ \\
	$R^{(2)}, D^{(2)} \leftarrow
		\begin{bsmallmatrix}
		    R^{(i)}_{(\mathcal{A})}(S) \\
		    \vdots \\
		    R^{(i)}_{(\mathcal{A})}(S)
		\end{bsmallmatrix},
		\begin{bsmallmatrix}
			D^{(i)}_{(\mathcal{A})}(S) \\
			\vdots \\
			D^{(i)}_{(\mathcal{A})}(S)
		\end{bsmallmatrix}$
	\STATE $\forall j=1, \ldots, n: \alpha(j)\leftarrow
		\begin{cases}
			1, & \textup{if $j=S$}.\\
			0, & \textup{otherwise}.
		\end{cases}$
	\STATE \textit{policy determination as policy iteration or linear program}\\
	       $\pi \leftarrow$ apply \eqref{eq:policy_iteration} to $\bar{P}^{(1)}$, $R^{(1)}$, $D^{(1)}$, $\bar{P}^{(2)}$, $R^{(2)}$, $D^{(2)}$, $\alpha$
	\STATE $\bar{P}_\pi$, $R_\pi$, $D_\pi$ $\leftarrow$ built from $\pi$ as defined in \eqref{eq:value_determination}
	\STATE $V_R \leftarrow (I-\bar{P}_\pi) V_R=R_\pi$ \\
	       $V_D \leftarrow (I-\bar{P}_\pi) V_D=D_\pi$
	\RETURN $(\alpha^T\cdot V_R)/(\alpha^T\cdot V_D)$
\end{algorithmic}

In this paper rationales are given but proofs must be deferred to an extended paper.  Conventional literature employs a stochastic matrix $P$ multiplied by a scalar discount factor $\beta$.  However, algorithm 1 can be operated with a more general {\em discounted transition matrix} $\bar{P}_{(\mathcal{A})}^{(i)}\doteq\textup{diag}(\beta^{(i)})P_{(\mathcal{A})}^{(i)}$ with $\beta^{(i)}$ bearing one discount factor per state, which can be interpreted as $1-\beta^{(i)}$ probability or terminating the process.  Rows 1 and 2 build the alternatives for continuing or restarting in state $S$ and $\alpha$ in row 3 is initialized to the current state, so that the subsequent policy iteration \eqref{eq:policy_iteration} in row 4 converges to the maximum {\em expected} ratio for evolutions from state $S$ onward.  The more computationally intensive \eqref{eq:lfp_as_lp} could be employed for more general cases. Row 6 obtains the total expected rewards from the optimal policy and row 7 returns the ratio of the rewards of state $S$ as fractional Gittins index.

A faster computation is given by algorithm 2 employing the elimination equivalence \cite{SONIN20081526}.
  Algorithm 2 can be operated with the same inputs as algorithm 1 and calculates all the Gittins indices of all states from the highest to the lowest.

\begin{algorithmic}[1]
	\stepcounter{algorithm}
	\renewcommand{\algorithmicensure}{\textbf{Algorithm 2:}}
	\renewcommand{\algorithmicrequire}{\textbf{Input:}}
	\ENSURE \textbf{Fractional Gittins Indices of all States}\\
	\textit{Calculation by state elimination}\label{alg:elimination}
	\REQUIRE
	\begin{tabular}{cp{17.5em}}
		$\bar{P}^{(i)}_{(\mathcal{A})}$ & $\forall x,y:\bar{P}^{(i)}_{(\mathcal{A})}(x,y)$ is the probability that process $i$ in state $x$ transitions to state $y$. \\
		$R^{(i)}_{(\mathcal{A})}$       & $\forall x: R^{(i)}_{(\mathcal{A})}(x)$ is the average information transported by process $i$ in state $x$. \\
		$D^{(i)}_{(\mathcal{A})}$       & $\forall x: D^{(i)}_{(\mathcal{A})}(x)$ is the average time process $i$ has sojourned in state $x$.
	\end{tabular}
	\STATE $P, R, D \leftarrow \bar{P}^{(i)}_{(\mathcal{A})}, R^{(i)}_{(\mathcal{A})}, D^{(i)}_{(\mathcal{A})}$\\
	       $G \in \mathbf{R}^n$,
	       $j \leftarrow \{1, \ldots, n\}$
	\WHILE {$|j|>0$}
	\STATE $q \leftarrow R/D$
	\STATE $s \leftarrow \argmax(q)$
	\STATE $c \leftarrow \{1,\ldots,|j|\}\setminus s$
	\STATE $G\left(j(s)\right) \leftarrow q(s)$
	\IF{$|c|>0$}
		\STATE $U \leftarrow P(x,y)|_{x\in c, y\in s}$ \\
		       $N \leftarrow (I-P(x,y)|_{x\in s, y\in s})^{-1}$\\
		       $T \leftarrow P(x,y)|_{x\in s, y\in c}$
		\STATE $P \leftarrow P(x,y)|_{x\in c, y\in c} + U N T$ \\
		       $R \leftarrow \left(R(x)|_{x\in c}^T + U N R|_{x\in s}^T\right)^T$ \\
		       $D \leftarrow \left(D(x)|_{x\in c}^T + U N D|_{x\in s}^T\right)^T$ \\
		       $j \leftarrow j\setminus s$
	\ENDIF
	\ENDWHILE
	\RETURN $G$
\end{algorithmic}

Algorithm 2 is described here in its most general form.  However, the matrix inversion shown in row 8 can be reduced to a simple division by letting row 4 return only one maximum of $q$ rather than all maxima altogether, i.e. $s$ has cardinality 1 ($|s|=1$).  The rationale behind algorithm 2 is that $R/D > R_\Sigma/D_\Sigma \rightarrow (R+R_\Sigma)/(D+D_\Sigma) > R_\Sigma/D_\Sigma$, i.e. if a transition leads to a state with higher immediate ratio, it is optimal {\em not} to stop because the overall ratio will be improved, hence only the maximum ratios in row 4 are optimal stopping states and can be eliminated for further calculations of lower indices. This is only true if the process is assumed to start in the current state and accrued rewards during the past evolution are discarded, which are exactly the assumptions for an optimality criterion explained in section \ref{sec:Optimality}.  This observation leads to an essential reuse of the algorithm in \cite{SONIN20081526} with the noteworthy feature that the termination probability in row 3 disappears because of the ratio maximization.

As a consistency check, it can be shown that the results returned by algorithms 1 and 2 are the same.

\section{Results}\label{sec:Results}

\begin{figure}
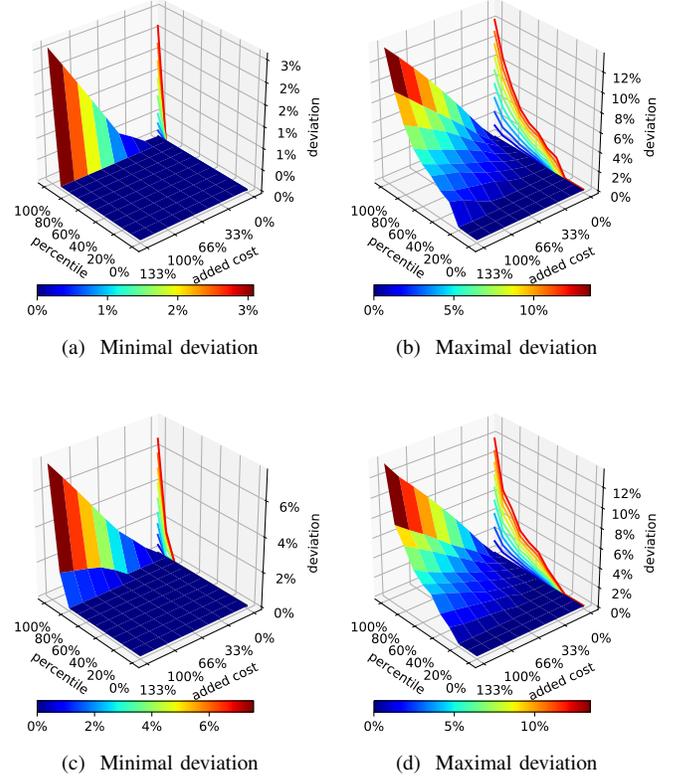

	\centering
	\subfloat[\label{fig:suboptimality_100_exp_0_percentile} Minimal deviation]{%
		{\footnotesize\includesvg[width=0.24\textwidth]{suboptimality_100_exp_2_arm_6_states_0_percentile}}
	}
	\subfloat[\label{fig:suboptimality_100_exp_100_percentile} Maximal deviation]{%
		{\footnotesize\includesvg[width=0.24\textwidth]{suboptimality_100_exp_2_arm_6_states_100_percentile}}
	}\\
	\subfloat[\label{fig:suboptimality_25_exp_0_percentile} Minimal deviation]{%
		{\footnotesize\includesvg[width=0.24\textwidth]{suboptimality_25_exp_3_arm_4_states_0_percentile}}
	}
	\subfloat[\label{fig:suboptimality_25_exp_100_percentile} Maximal deviation]{%
		{\footnotesize\includesvg[width=0.24\textwidth]{suboptimality_25_exp_3_arm_4_states_100_percentile}}
	}
	\caption{Deviations for the 100 6-states/2-alternatives models in figures \ref{fig:suboptimality_100_exp_0_percentile} and \ref{fig:suboptimality_100_exp_100_percentile} and for the 25 4-states/3-alternatives models in figures \ref{fig:suboptimality_25_exp_0_percentile} and \ref{fig:suboptimality_25_exp_100_percentile}.}
	\label{fig:suboptimality_0_100_percentiles}
\end{figure}

\begin{figure}
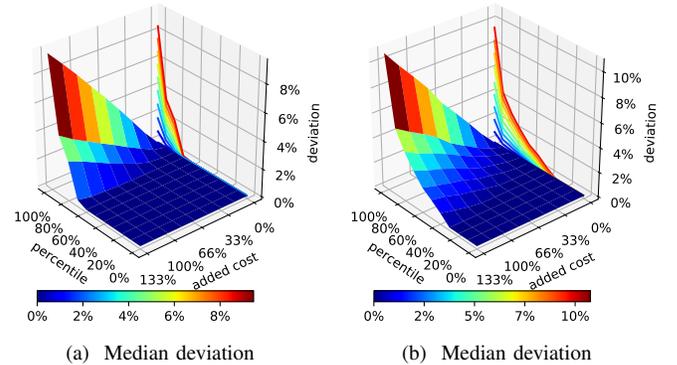
 
	\centering
	\subfloat[\label{fig:suboptimality_100_exp_50_percentile} Median deviation]{%
		{\footnotesize\includesvg[width=0.24\textwidth]{suboptimality_100_exp_2_arm_6_states_50_percentile}}
	}
	\subfloat[\label{fig:suboptimality_25_exp_50_percentile} Median deviation]{%
		{\footnotesize\includesvg[width=0.24\textwidth]{suboptimality_25_exp_3_arm_4_states_50_percentile}}
	}
	\caption{Deviations for the 100 6-states/2-alternatives models in figure \ref{fig:suboptimality_100_exp_50_percentile} and for the 25 4-states/3-alternatives models in figure \ref{fig:suboptimality_25_exp_50_percentile}.}
	\label{fig:suboptimality_50_percentiles}
\end{figure}

The index policy induced by the fractional Gittins indices becomes suboptimal for the {\em restless bandit}, or in simple words when the alternative not selected keeps evolving.  In this case the optimal policy must be calculated by \eqref{eq:policy_iteration} or \eqref{eq:lfp_as_lp} for the Markov decision process built by \eqref{eq:bandit_as_mdp} and this is feasible in general only for small space and action states (but see \cite{ninomora2001}).

For a numerical analysis, 100 small random models of 6 states with 2 alternatives and 25 ones of 4 states with 3 alternatives have been generated with characteristics that mimic handover procedures.  They bear a cost for switching between alternatives as additional delay ranging from 0\% to 133\% of the average sojourn time in a handover state.  Both the average probabilities of handover failure and of connection loss initiating a new handover are at 5\% (frequent handover).

The {\em deviation} of the index policy against the optimal policy is $(r_\textup{opt} - r_{\textup{index}})/r_\textup{opt}$, where $r_\textup{opt}$ is the ratio obtained from the optimal policy calculated on \eqref{eq:bandit_as_mdp} as {\em restless bandit} by \eqref{eq:policy_iteration} and $r_\textup{index}$ is the ratio obtained from the policy calculated by algorithm 2 on every $\{\bar{P}_{(\mathcal(A))}^{(i)}, R_{(\mathcal(A))}^{(i)}, D_{(\mathcal(A))}^{(i)}\}$ individually, as {\em restful bandit} assuming $\bar{P}_{(\mathcal(P))}^{(i)}=I$ and $R_{(\mathcal(P))}^{(i)}=\vec{0}$.

Figure \ref{fig:suboptimality_0_100_percentiles} represents the distribution of the maximal and minimal deviations across models when cost ranges from 0\% to 133\%.  A {\em percentile} just below 100\% shows the highest deviation, one just above 0\% shows the minimal one.  Figure \ref{fig:suboptimality_50_percentiles} provides the same representation for median deviations.

The highest deviation at 12\% implies that no more than 12\% of the theoretically achievable maximal bandwidth is missed by the fractional Gittins index policy.  The median deviations never exceed 10\%, but already the median of highest deviations across all handovers (50\% percentile in figure \ref{fig:suboptimality_25_exp_100_percentile}) does not exceed 2\%.  Median deviations across handovers are ``typically'' (50\% percentile in figure \ref{fig:suboptimality_25_exp_50_percentile}, representing half of the deviations for half of the handovers) below 0.5\%.

\section{Conclusion}

This work has advanced the theory of Gittins indices to the fractional case in the Markov ratio decision processes and has employed it to define bandwidth-optimized handovers as {\em multiarmed bandits}.  Although the determination of the optimal policy in the most general case is prohibitively (PSPACE) complex, the novel fractional Gittins indices provide a simple optimal solution for the {\em restful bandit} and an effective one for the most general case of the {\em restless bandit}.

A numerical analysis on a wide range of models has shown that in the general case the index policy induced by novel fractional Gittins indices never misses more than 12\% of the theoretically achievable bandwidth (from 2\% to 0.5\% in typical cases): although these models must be small due to the complexity of computing their theoretical optimum, the analysis confirms and quantifies the bound on optimality.

The framework introduced in this paper has shown the limits of current sticky handover policies like CRE and how they can be redefined for scaling up to large heterogeneous networks through properly designed dynamical priorities.  While bandwidth is surely not the sole optimization criteria for handovers, this work provides a framework to devise optimal stopping for fractional Markovian multi-objective selection procedures as employed in handovers.  Although rationales for proofs have been given in this paper, an extended paper is due to give proofs and further characterize the reach of the theory.

{\em Acknowledgment}: Part of the research in this work has received funding from the European Commission H2020 programme, grant agreement No 815178 (5GENESIS project).




\bibliographystyle{IEEEtran}
\bibliography{IEEEabrv,literature}
%
%
%

\end{document}